\documentclass[conference,letterpaper]{IEEEtran}

\usepackage[utf8]{inputenc} 
\usepackage[T1]{fontenc}
\usepackage{url}
\usepackage{ifthen}
\usepackage{cite}
\usepackage[cmex10]{amsmath} 

\usepackage{hyperref}%
\usepackage{amsfonts}
\usepackage{amsmath}
\usepackage{amssymb}

\usepackage{mathtools}%

\newtheorem{theorem}{Theorem}
\newtheorem{remark}{Remark}
\newtheorem{proposition}{Proposition}
\newtheorem{corollary}{Corollary}

\begin{document}
\title{Second Law of Entanglement Dynamics for the Non-Asymptotic Regime} 


\author{%
\IEEEauthorblockN{Mark M. Wilde}
\IEEEauthorblockA
{Hearne Institute for Theoretical Physics, Department of Physics and Astronomy, and Center for Computation and Technology\\Louisiana State University, Baton Rouge, Louisiana 70803, USA, 
Email: mwilde@lsu.edu}}


\maketitle

\begin{abstract}
  The distillable entanglement of a bipartite quantum state does not exceed its
entanglement cost. This well known inequality can be understood as a second
law of entanglement dynamics in the asymptotic regime of entanglement
manipulation, excluding the possibility of perpetual entanglement extraction
machines that generate boundless entanglement from a finite reserve. In this
paper, I establish a refined second law of entanglement dynamics that holds for the
non-asymptotic regime of entanglement manipulation.
\end{abstract}


\section{Introduction}

Entanglement is a fundamental resource for quantum information processing, as
it is the enabling fuel for critical protocols like teleportation
\cite{BBC+93}, super-dense coding \cite{PhysRevLett.69.2881}, and quantum key
distribution \cite{E91}. As such, it has been a longstanding challenge to
understand entanglement as a resource and to quantify it
\cite{BBPSSW96EPP,BDSW96}, and this subject is known as entanglement theory
(see \cite{HHHH09,PV07,W18thesis,KW20book} for reviews of the topic, as well
as the latest results).

Two basic operational quantities of interest in entanglement theory are the
distillable entanglement and the entanglement cost of a bipartite state
$\rho_{AB}$ \cite{BDSW96,HHT01}. The physical scenario corresponding to these
quantities is that Alice and Bob are in distant laboratories, a third party
distributes system $A$ of $\rho_{AB}$ to Alice and system $B$ of $\rho_{AB}$
to Bob, and they are allowed to perform local operations and classical
communication (LOCC) on this state. The distillable entanglement is defined to
be the maximal rate at which ebits (Bell states) can be extracted from a large
number $n$\ of copies of $\rho_{AB}$ by means of an entanglement distillation
protocol, i.e., when using LOCC\ for free and such that the fidelity of the
actual output state to the desired ideal ebits approaches one in the limit
$n\rightarrow\infty$. The entanglement cost is defined to be the minimal rate
at which ebits are needed to generate a large number $n$ of copies of
$\rho_{AB}$ by means of an entanglement dilution protocol, i.e., when using
LOCC\ for free and such that the fidelity of the actual output to the ideal
state $\rho_{AB}^{\otimes n}$ approaches one in the limit $n\rightarrow\infty
$. Both the distillable entanglement and the entanglement cost are notoriously
difficult to calculate in general, and it is even suspected that these
quantities are uncomputable in the Turing sense \cite{WCP11}.

It has long been understood that the distillable entanglement does not exceed
the entanglement cost \cite{BDSW96,PhysRevA.65.012323}. This inequality can be
interpreted as a ``second law of entanglement dynamics,'' preventing the
existence of perpetual entanglement extraction devices that generate an
unbounded amount of entanglement from a finite reserve. The inequality indeed
follows from basic reasoning akin to that for the second law of thermodynamics
and against perpetual motion machines:\ If the inequality were not to hold,
then it would be possible to produce a boundless amount of entanglement, by
repeatedly executing a protocol for entanglement distillation followed by one
for entanglement dilution. This is intuitively impossible, and so the
distillable entanglement cannot exceed the entanglement cost. See \cite{Wat18} for a formal proof and \cite{BP08,BP2010} for a strengthened second law of entanglement dynamics that holds for free operations beyond LOCC.

The reasoning given above applies in the asymptotic regime of a large number
$n$ of copies of the state $\rho_{AB}$ and with fidelities tending to one in
the limit $n\rightarrow\infty$. However, this reasoning does not apply in the
non-asymptotic regime \cite{BD10a,BD11} of interest for practical applications
and near-term quantum devices. As such, we are left to wonder what kind of
relationship might hold in the non-asymptotic regime; i.e., what is a second
law of entanglement dynamics for the non-asymptotic regime?

In this paper, I establish a fundamental inequality relating distillable
entanglement and entanglement cost in the non-asymptotic regime (see
Theorem~\ref{thm:main-result}), which addresses the aforementioned question.
This inequality states that the one-shot distillable entanglement does
not exceed the one-shot entanglement cost plus an additional finite-size
correction term that depends on the errors of the transformations
corresponding to distillation and dilution. In the regime in which the errors
are small, this correction term is approximately linear in the total error,
indicating that the one-shot distillable entanglement cannot be much
larger than the one-shot entanglement cost. However, when the errors are
large (i.e., near to one), the correction term can be rather large, so that
the inequality is not particularly relevant. Furthermore, the asymptotic
statement mentioned above is recovered from Theorem~\ref{thm:main-result} by
applying limits and the definitions of distillable entanglement and
entanglement cost in the asymptotic regime.

The rest of this paper proceeds by establishing notation and defining the one-shot distillable
entanglement and one-shot entanglement cost of a bipartite state. I then prove the main result (Theorem~\ref{thm:main-result}) in three parts. First, I recall that the $\varepsilon$-Rains relative
entropy \cite{TWW17,TBR15} is an upper bound on the one-shot distillable
entanglement \cite{FWTD19,KW20book}. The second part is the main technical result: an exact
evaluation of the $\varepsilon$-Rains relative entropy of a maximally
entangled state. The third part consists of applying this identity in the analysis of a quasi-cyclic process
that dilutes a maximally entangled state to a generic bipartite state and then
distills that back to another maximally entangled state. After that, I show how to recover the asymptotic statement of the second law by taking limits, and I establish an alternate non-asymptotic second law when errors are measured with normalized trace distance rather than fidelity. Finally, I discuss how this inequality extends much more
generally to the entanglement theory of bipartite quantum channels
\cite{BHLS03,BDW18,DBW20,BDWW19,GS19}.

\section{Definitions}

Nearly all concepts discussed in this preliminary section are reviewed in
detail in \cite{KW20book}. Let us begin by defining $\Phi_{AB}^{d}$ as the
following maximally entangled state of Schmidt rank $d$:%
\begin{equation}
\Phi_{AB}^{d}\coloneqq \frac{1}{d}\sum_{i,j=0}^{d-1}|i\rangle\!\langle j|_{A}%
\otimes|i\rangle\!\langle j|_{B},
\end{equation}
where $\{|i\rangle_{A}\}_{i}$ and $\{|i\rangle_{B}\}_{i}$ are orthonormal bases.

A bipartite channel is an LOCC channel if it can be realized as a finite, yet
arbitrarily large number of compositions of one-way LOCC channels of the
following form \cite{CLM+14,KW20book}:%
\begin{align}
\mathcal{L}_{AB\rightarrow A^{\prime}B^{\prime}}^{\rightarrow}  &  =\sum
_{x}\mathcal{E}_{A\rightarrow A^{\prime}}^{x}\otimes\mathcal{F}_{B\rightarrow
B^{\prime}}^{x},\\
\mathcal{L}_{AB\rightarrow A^{\prime}B^{\prime}}^{\leftarrow}  &  =\sum
_{x}\mathcal{G}_{A\rightarrow A^{\prime}}^{x}\otimes\mathcal{K}_{B\rightarrow
B^{\prime}}^{x},
\end{align}
where $\{\mathcal{E}_{A\rightarrow A^{\prime}}^{x}\}_{x}$ and $\{\mathcal{K}%
_{B\rightarrow B^{\prime}}^{x}\}_{x}$ are sets of completely positive maps
such that the sum maps $\sum_{x}\mathcal{E}_{A\rightarrow A^{\prime}}^{x}$ and
$\sum_{x}\mathcal{K}_{B\rightarrow B^{\prime}}^{x}$ are trace preserving and
$\{\mathcal{F}_{B\rightarrow B^{\prime}}^{x}\}_{x}$ and $\{\mathcal{G}%
_{A\rightarrow A^{\prime}}^{x}\}_{x}$ are sets of quantum channels (i.e.,
completely positive, trace-preserving maps). An LOCC channel $\mathcal{L}%
_{AB\rightarrow A^{\prime}B^{\prime}}$ can be written in the following
separable form:%
\begin{equation}
\mathcal{L}_{AB\rightarrow A^{\prime}B^{\prime}}=\sum_{y}\mathcal{P}%
_{A\rightarrow A^{\prime}}^{y}\otimes\mathcal{Q}_{B\rightarrow B^{\prime}}%
^{y}, \label{eq:LOCC-channel}%
\end{equation}
where $\{\mathcal{P}_{A\rightarrow A^{\prime}}^{y}\}_{y}$ and $\{\mathcal{Q}%
_{B\rightarrow B^{\prime}}^{y}\}_{y}$ are sets of completely positive maps
such that $\mathcal{L}_{AB\rightarrow A^{\prime}B^{\prime}}$ is trace
preserving. However, the converse statement is not true
\cite{PhysRevA.59.1070}; i.e., not every channel that can be written as in
\eqref{eq:LOCC-channel}\ can be realized by LOCC.

The fidelity of quantum states $\omega$ and $\tau$ is defined as
$F(\omega,\tau)\coloneqq \left\Vert \sqrt{\omega}\sqrt{\tau}\right\Vert _{1}^{2}$
\cite{Uhl76}, and the trace distance as $\left\Vert \omega-\tau\right\Vert
_{1}$, where $\left\Vert A\right\Vert _{1}=\operatorname{Tr}[\sqrt{A^{\dag}%
A}]$ is the trace norm (i.e., Schatten 1-norm). The sine distance of $\omega$
and $\tau$ is defined as $P(\omega,\tau)\coloneqq \sqrt{1-F(\omega,\tau)}$
\cite{R02,R03,GLN04,R06}, and it obeys the triangle inequality, as well as the
data-processing inequality $P(\omega,\tau)\geq P(\mathcal{N}(\omega
),\mathcal{N}(\tau))$, where $\mathcal{N}$ is a quantum channel.

The one-shot distillable entanglement $E_{D}^{\varepsilon}(A;B)_{\rho}$ of a
bipartite state $\rho_{AB}$ is defined for $\varepsilon\in\left[  0,1\right]
$ as \cite{BD10a}%
\begin{multline}
E_{D}^{\varepsilon}(A;B)_{\rho}\coloneqq \label{def:one-shot-ent-dist}\\
\sup_{\substack{d\in\mathbb{N},\\\mathcal{L}\in\text{LOCC}}}\left\{  \log
_{2}d:F(\mathcal{L}_{AB\rightarrow\hat{A}\hat{B}}(\rho_{AB}),\Phi_{\hat{A}%
\hat{B}}^{d})\geq1-\varepsilon\right\}  .
\end{multline}
In words, it is equal to the maximum number of $\varepsilon$-approximate ebits
that one can distill from $\rho_{AB}$ by means of LOCC. The one-shot
entanglement cost of $\rho_{AB}$ is defined for $\varepsilon\in\left[
0,1\right]  $ as \cite{BD11}%
\begin{multline}
E_{C}^{\varepsilon}(A;B)_{\rho}\coloneqq \label{def:one-shot-ent-cost}\\
\inf_{\substack{d\in\mathbb{N},\\\mathcal{L}\in\text{LOCC}}}\left\{  \log
_{2}d:F(\mathcal{L}_{\hat{A}\hat{B}\rightarrow AB}(\Phi_{\hat{A}\hat{B}}%
^{d}),\rho_{AB})\geq1-\varepsilon\right\}  .
\end{multline}
In words, it is equal to the minimum number of ebits that is required to
generate $\rho_{AB}$ approximately by means of LOCC. Observe that the function
$\varepsilon\rightarrow E_{D}^{\varepsilon}(A;B)_{\rho}$ is monotone
non-decreasing while the function $\varepsilon\rightarrow E_{C}^{\varepsilon
}(A;B)_{\rho}$ is monotone non-increasing. One can alternatively define the
approximation error for $E_{D}^{\varepsilon}(A;B)_{\rho}$ and $E_{C}%
^{\varepsilon}(A;B)_{\rho}$\ in terms of normalized trace distance instead of
fidelity, which we consider later on in Section~\ref{sec:trace-dist-err}.

\section{Second law of entanglement dynamics in the non-asymptotic regime}

The main result of this paper is the following inequality that relates
distillable entanglement and entanglement cost in the non-asymptotic regime,
interpreted as a second law of entanglement dynamics:

\begin{theorem}
\label{thm:main-result}Let $\rho_{AB}$ be a bipartite state, and let
$\varepsilon_{1},\varepsilon_{2}\in\left[  0,1\right]  $ be such that
$\varepsilon^{\prime}\coloneqq  \left[  \sqrt{\varepsilon_{1}}+\sqrt
{\varepsilon_{2}}\right]  ^{2}<1$. Then%
\begin{equation}
E_{D}^{\varepsilon_{2}}(A;B)_{\rho}\leq E_{C}^{\varepsilon_{1}}(A;B)_{\rho
}+\log_{2}\!\left(  \frac{1}{1-\varepsilon^{\prime}}\right)  
\label{eq:cost-and-distill}.
\end{equation}
\end{theorem}

\begin{remark}
Given that $\log_{2}\!\left(  \frac{1}{1-x}\right)  =x/\ln2+O(x^{2})$ for
$x\approx0$, the inequality in \eqref{eq:cost-and-distill}\ asserts that the
one-shot distillable entanglement is bounded from above by the one-shot
entanglement cost plus a small correction term when the error sum
$\varepsilon^{\prime}$\ is small. When the error sum $\varepsilon^{\prime}$ is
large (i.e., $\approx1$), the inequality is loose and it does not exclude the
possibility of the one-shot distillable entanglement exceeding the one-shot
entanglement cost.
\end{remark}

\medskip

\begin{IEEEproof}
[Proof of Theorem~\ref{thm:main-result}]The proof involves three parts. First,
let us recall that the one-shot distillable entanglement $E_{D}^{\varepsilon
}(A;B)_{\rho}$ is bounded from above by the $\varepsilon$-Rains relative
entropy for all $\varepsilon\in\left[  0,1\right]  $:%
\begin{equation}
E_{D}^{\varepsilon}(A;B)_{\rho}\leq R_{H}^{\varepsilon}(A;B)_{\rho
},\label{eq:one-shot-dist-ent-rains-rel-ent}%
\end{equation}
where the $\varepsilon$-Rains relative entropy is defined as
\cite{TWW17,TBR15}%
\begin{equation}
R_{H}^{\varepsilon}(A;B)_{\rho}\coloneqq \min_{\sigma_{AB}\in\text{PPT}^{\prime}%
(A:B)}D_{H}^{\varepsilon}(\rho_{AB}\Vert\sigma_{AB}),
\end{equation}
the hypothesis testing relative entropy $D_{H}^{\varepsilon}(\omega\Vert\tau
)$\ is defined for a state $\omega$ and a positive semi-definite operator
$\tau$ as \cite{BD10,BD11ieee,WR12}%
\begin{equation}
D_{H}^{\varepsilon}(\omega\Vert\tau)\coloneqq -\log_{2}\left\{  \min_{\Lambda\geq
0}\operatorname{Tr}[\Lambda\tau]:\operatorname{Tr}[\Lambda\omega
]\geq1-\varepsilon,\Lambda\leq I\right\}  ,
\end{equation}
and the set PPT$^{\prime}(A:B)$ as \cite{AdMVW02}%
\begin{equation}
\text{PPT}^{\prime}(A:B)\coloneqq \left\{  \sigma_{AB}:\sigma_{AB}\geq0,\left\Vert
T_{B}(\sigma_{AB})\right\Vert _{1}\leq1\right\}  .
\end{equation}
Note that \cite{KW17a,KW20book}%
\begin{equation}
D_{H}^{\varepsilon}(\omega\Vert\tau)\coloneqq -\log_{2}\left\{  \min_{\Lambda\geq
0}\operatorname{Tr}[\Lambda\tau]:\operatorname{Tr}[\Lambda\omega
]=1-\varepsilon,\Lambda\leq I\right\}  .
\end{equation}
In the above, $T_{B}$ denotes the partial transpose. The inequality in
\eqref{eq:one-shot-dist-ent-rains-rel-ent} follows as a consequence of
Theorem~4\ of \cite{FWTD19}. Alternatively, see Theorem~8.7 of \cite{KW20book}.

The next step involves the following identity for the $\varepsilon$-Rains
relative entropy of a maximally entangled state $\Phi_{AB}^{d}$:%
\begin{equation}
R_{H}^{\varepsilon}(A;B)_{\Phi^{d}}=\log_{2}d+\log_{2}\!\left(  \frac
{1}{1-\varepsilon}\right)  . \label{eq:eps-Rains-identity-max-ent}%
\end{equation}
This identity is established in
Proposition~\ref{prop:Rains-ident-max-ent-state} and is the main technical
contribution of this paper.

Finally, let us consider the following sequence of transformations:%
\begin{equation}
\Phi_{A^{\prime}B^{\prime}}^{d_{\text{in}}}\quad\underrightarrow
{\varepsilon_{1}}\quad\rho_{AB}\quad\underrightarrow{\varepsilon_{2}}\quad
\Phi_{A^{\prime\prime}B^{\prime\prime}}^{d_{\text{out}}},
\label{eq:main-argument-conceptual}%
\end{equation}
where the arrows indicate that the transformations take place by means of LOCC
channels\ $\mathcal{L}_{A^{\prime}B^{\prime}\rightarrow AB}^{1}$ and
$\mathcal{L}_{AB\rightarrow A^{\prime\prime}B^{\prime\prime}}^{2}$, with an
approximation error of $\varepsilon_{1}$ and $\varepsilon_{2}$, respectively.
By making use of the definition of and the triangle inequality for the sine
distance and its data-processing inequality, the total error of the
transformation from $\Phi_{A^{\prime}B^{\prime}}^{d_{\text{in}}}$ to
$\Phi_{A^{\prime\prime}B^{\prime\prime}}^{d_{\text{out}}}$ is no larger than
$\varepsilon^{\prime}$ because the following inequalities hold by assumption%
\begin{align}
F(\mathcal{L}_{A^{\prime}B^{\prime}\rightarrow AB}^{1}(\Phi_{A^{\prime
}B^{\prime}}^{d_{\text{in}}}),\rho_{AB})  &  \geq1-\varepsilon_{1},\\
F(\mathcal{L}_{AB\rightarrow A^{\prime\prime}B^{\prime\prime}}^{2}(\rho
_{AB}),\Phi_{A^{\prime\prime}B^{\prime\prime}}^{d_{\text{out}}})  &
\geq1-\varepsilon_{2},
\end{align}
so that%
\begin{equation}
P((\mathcal{L}_{AB\rightarrow A^{\prime\prime}B^{\prime\prime}}^{2}%
\circ\mathcal{L}_{A^{\prime}B^{\prime}\rightarrow AB}^{1})(\Phi_{A^{\prime
}B^{\prime}}^{d_{\text{in}}}),\Phi_{A^{\prime\prime}B^{\prime\prime}%
}^{d_{\text{out}}})\leq\sqrt{\varepsilon_{1}}+\sqrt{\varepsilon_{2}}.
\end{equation}
The transformation in \eqref{eq:main-argument-conceptual} can be understood as
a particular way to perform entanglement distillation of the state
$\Phi_{A^{\prime}B^{\prime}}^{d_{\text{in}}}$ to the state $\Phi
_{A^{\prime\prime}B^{\prime\prime}}^{d_{\text{out}}}$ with error
$\varepsilon^{\prime}$. As such, we find that%
\begin{equation}
\log_{2}d_{\text{out}}\leq\log_{2}d_{\text{in}}+\log_{2}\!\left(  \frac
{1}{1-\varepsilon^{\prime}}\right)  , \label{eq:main-bound-cyclic-proc}%
\end{equation}
because%
\begin{align}
\log_{2}d_{\text{out}}  &  \leq E_{D}^{\varepsilon^{\prime}}(A^{\prime
};B^{\prime})_{\Phi^{d_{\text{in}}}}\\
&  \leq R_{H}^{\varepsilon^{\prime}}(A^{\prime};B^{\prime})_{\Phi
^{d_{\text{in}}}}\\
&  =\log_{2}d_{\text{in}}+\log_{2}\!\left(  \frac{1}{1-\varepsilon^{\prime}%
}\right)  .
\end{align}
The first inequality is a consequence of the definition of one-shot
distillable entanglement in \eqref{def:one-shot-ent-dist}. The second
inequality follows from \eqref{eq:one-shot-dist-ent-rains-rel-ent}, and the
equality follows from \eqref{eq:eps-Rains-identity-max-ent}. Since the
inequality in \eqref{eq:main-bound-cyclic-proc} holds for every entanglement
dilution protocol $\mathcal{L}_{A^{\prime}B^{\prime}\rightarrow AB}^{1}$
taking $\Phi_{A^{\prime}B^{\prime}}^{d_{\text{in}}}$ to $\rho_{AB}$ with error
$\varepsilon_{1}$ and for every entanglement distillation protocol
$\mathcal{L}_{AB\rightarrow A^{\prime\prime}B^{\prime\prime}}^{2}$ taking
$\rho_{AB}$ to $\Phi_{A^{\prime\prime}B^{\prime\prime}}^{d_{\text{out}}}$ with
error $\varepsilon_{2}$, we can take an infimum over $d_{\text{in}}$ and a
supremum over $d_{\text{out}}$, apply the definitions in
\eqref{def:one-shot-ent-cost} and \eqref{def:one-shot-ent-dist}, respectively,
and conclude the inequality in \eqref{eq:cost-and-distill}.
\end{IEEEproof}

\medskip

Let us now prove \eqref{eq:eps-Rains-identity-max-ent}:

\medskip

\begin{proposition}
\label{prop:Rains-ident-max-ent-state} For $\varepsilon\in\lbrack0,1)$, the
$\varepsilon$-Rains relative entropy of the maximally entangled state
$\Phi_{AB}^{d}$ is as follows:%
\begin{equation}
R_{H}^{\varepsilon}(A;B)_{\Phi^{d}}=\log_{2}d+\log_{2}\!\left(  \frac
{1}{1-\varepsilon}\right)  .
\end{equation}

\end{proposition}

\begin{IEEEproof}
The maximally entangled state $\Phi_{AB}^{d}$ is invariant under a bilateral
twirl:%
\begin{equation}
\Phi_{AB}^{d}=\mathcal{T}_{AB}(\Phi_{AB}^{d}),
\end{equation}
where%
\begin{equation}
\mathcal{T}_{AB}(X_{AB})\coloneqq \int dU\ \left(  U_{A}\otimes\overline{U}%
_{B}\right)  X_{AB}\left(  U_{A}\otimes\overline{U}_{B}\right)  ^{\dag}.
\end{equation}
Recall that \cite{Wat18}%
\begin{multline}
\mathcal{T}_{AB}(X_{AB})=\Phi_{AB}\operatorname{Tr}[\Phi_{AB}X_{AB}%
]\label{eq:result-of-twirl}\\
+\frac{I_{AB}-\Phi_{AB}}{d^{2}-1}\operatorname{Tr}[(I_{AB}-\Phi_{AB})X_{AB}].
\end{multline}
The twirling channel $\mathcal{T}_{AB}$ is an LOCC\ channel. As such, for
every operator $\sigma_{AB}\in\ $PPT$^{\prime}(A\!:\!B)$, it follows that $\mathcal{T}_{AB}(\sigma_{AB})\in\ $PPT$^{\prime
}(A:B)$ \cite{R99,R01,TWW17}, and we find that%
\begin{align}
D_{H}^{\varepsilon}(\Phi_{AB}\Vert\sigma_{AB}) &  \geq D_{H}^{\varepsilon
}(\mathcal{T}_{AB}(\Phi_{AB})\Vert\mathcal{T}_{AB}(\sigma_{AB}))\\
&  =D_{H}^{\varepsilon}(\Phi_{AB}\Vert\mathcal{T}_{AB}(\sigma_{AB})),
\end{align}
where we used the data-processing inequality for the hypothesis testing
relative entropy. Thus, it suffices to minimize $R_{H}^{\varepsilon
}(A;B)_{\Phi^{d}}$ with respect to $\mathcal{T}_{AB}(\sigma_{AB})$. By
applying \eqref{eq:result-of-twirl}, it follows that all such states have the
following form:%
\begin{equation}
\mathcal{T}_{AB}(\sigma_{AB})=\alpha\Phi_{AB}+\beta\left(  \frac{I_{AB}%
-\Phi_{AB}}{d^{2}-1}\right)  ,
\end{equation}
where $\alpha,\beta\in\left[  0,1\right]  $ are such that $\mathcal{T}%
_{AB}(\sigma_{AB})\in\ $PPT$^{\prime}(A:B)$. We now determine the conditions
on $\alpha$ and $\beta$ such that $\mathcal{T}_{AB}(\sigma_{AB})\in
\ $PPT$^{\prime}(A:B)$. We first require $\alpha,\beta\geq0$ so that
$\mathcal{T}_{AB}(\sigma_{AB})$ is positive semi-definite. Also, consider that%
\begin{align}
&  \left\Vert \alpha T_{B}(\Phi_{AB})+\beta T_{B}\!\left(  \frac{I_{AB}%
-\Phi_{AB}}{d^{2}-1}\right)  \right\Vert _{1}\nonumber\\
&  =\left\Vert \frac{\alpha}{d}F_{AB}+\beta\left(  \frac{I_{AB}-\frac{1}%
{d}F_{AB}}{d^{2}-1}\right)  \right\Vert _{1}\\
&  =\left\Vert \frac{1}{d}\left(  \alpha-\frac{\beta}{d^{2}-1}\right)
F_{AB}+\frac{\beta}{d^{2}-1}I_{AB}\right\Vert _{1}%
,\label{eq:plug-in-proof-step}%
\end{align}
where we applied the fact that%
\begin{equation}
T_{B}(\Phi_{AB})=\frac{1}{d}F_{AB},
\end{equation}
with $F_{AB}$ the unitary swap operator:%
\begin{equation}
F_{AB}\coloneqq \sum_{i,j=0}^{d-1}|i\rangle\!\langle j|_{A}\otimes|j\rangle\!\langle
i|_{B}.
\end{equation}
Now consider defining the projections $\Pi_{AB}^{\mathcal{S}}$ and $\Pi
_{AB}^{\mathcal{A}}$ onto the symmetric and antisymmetric subspaces,
respectively, in terms of $I_{AB}=\Pi_{AB}^{\mathcal{S}}+\Pi_{AB}%
^{\mathcal{A}}$ and $F_{AB}=\Pi_{AB}^{\mathcal{S}}-\Pi_{AB}^{\mathcal{A}}$.
Plugging in to \eqref{eq:plug-in-proof-step}, we find that%
\begin{align}
&  \left\Vert
\begin{array}
[c]{c}%
\frac{1}{d}\left(  \alpha-\frac{\beta}{d^{2}-1}\right)  \left(  \Pi
_{AB}^{\mathcal{S}}-\Pi_{AB}^{\mathcal{A}}\right)  \\
+\frac{\beta}{d^{2}-1}\left(  \Pi_{AB}^{\mathcal{S}}+\Pi_{AB}^{\mathcal{A}%
}\right)
\end{array}
\right\Vert _{1}\nonumber\\
&  =\left\Vert
\begin{array}
[c]{c}%
\left(  \frac{1}{d}\left(  \alpha-\frac{\beta}{d^{2}-1}\right)  +\frac{\beta
}{d^{2}-1}\right)  \Pi_{AB}^{\mathcal{S}}\\
+\left(  \frac{\beta}{d^{2}-1}-\frac{1}{d}\left(  \alpha-\frac{\beta}{d^{2}%
-1}\right)  \right)  \Pi_{AB}^{\mathcal{A}}%
\end{array}
\right\Vert _{1}\\
&  =\left\vert \frac{1}{d}\left(  \alpha-\frac{\beta}{d^{2}-1}\right)
+\frac{\beta}{d^{2}-1}\right\vert \operatorname{Tr}[\Pi_{AB}^{\mathcal{S}%
}]\nonumber\\
&  \qquad+\left\vert \frac{\beta}{d^{2}-1}-\frac{1}{d}\left(  \alpha
-\frac{\beta}{d^{2}-1}\right)  \right\vert \operatorname{Tr}[\Pi
_{AB}^{\mathcal{A}}]\\
&  =\left\vert \frac{1}{d}\left(  \alpha-\frac{\beta}{d^{2}-1}\right)
+\frac{\beta}{d^{2}-1}\right\vert \frac{d\left(  d+1\right)  }{2}\nonumber\\
&  \qquad+\left\vert \frac{\beta}{d^{2}-1}-\frac{1}{d}\left(  \alpha
-\frac{\beta}{d^{2}-1}\right)  \right\vert \frac{d\left(  d-1\right)  }{2}.
\end{align}
Continuing, the last line above is equal to%
\begin{align}
&  \left\vert \alpha-\frac{\beta}{d^{2}-1}+\frac{d\beta}{d^{2}-1}\right\vert
\frac{d+1}{2}\nonumber\\
&  \qquad+\left\vert \frac{d\beta}{d^{2}-1}-\left(  \alpha-\frac{\beta}%
{d^{2}-1}\right)  \right\vert \frac{d-1}{2}\nonumber\\
&  =\left\vert \alpha+\frac{\left(  d-1\right)  \beta}{d^{2}-1}\right\vert
\frac{d+1}{2}+\left\vert \frac{\left(  d+1\right)  \beta}{d^{2}-1}%
-\alpha\right\vert \frac{d-1}{2}\\
&  =\left\vert \alpha+\frac{\beta}{d+1}\right\vert \frac{d+1}{2}+\left\vert
\frac{\beta}{d-1}-\alpha\right\vert \frac{d-1}{2}\\
&  =\frac{1}{2}\left[  \alpha\left(  d+1\right)  +\beta+\left\vert
\beta-\alpha\left(  d-1\right)  \right\vert \right]  \\
&  =\left\{
\begin{array}
[c]{cc}%
\alpha+\beta & \beta\geq\alpha\left(  d-1\right)  \\
\alpha d & \beta<\alpha\left(  d-1\right)
\end{array}
\right.  .
\end{align}
Thus, to have that $\mathcal{T}_{AB}(\sigma_{AB})\in\ $PPT$^{\prime}(A:B)$, we
require that $\alpha,\beta\geq0$ and $\alpha+\beta\leq1$ if $\beta\geq
\alpha\left(  d-1\right)  $ and $\alpha d\leq1$ if $\beta<\alpha\left(
d-1\right)  $. Let PPT$^{\prime}$ be a shorthand for the set of $\alpha$ and
$\beta$ satisfying these conditions. Note that $\beta\geq\alpha\left(
d-1\right)  $ implies that%
\begin{equation}
1\geq\alpha+\beta\geq\alpha+\alpha\left(  d-1\right)  =\alpha d,
\end{equation}
so that%
\begin{equation}
\alpha\leq\frac{1}{d}\quad\text{for all~}(\alpha,\beta)\in\text{PPT}^{\prime}.
\end{equation}
Then we find that%
\begin{align}
&  \inf_{\sigma_{AB}\in\text{PPT}^{\prime}(A:B)}D_{H}^{\varepsilon}(\rho
_{AB}\Vert\sigma_{AB})\nonumber\\
&  =\inf_{(\alpha,\beta)\in\text{PPT}^{\prime}}D_{H}^{\varepsilon}\!\left(
\Phi_{AB}\middle\Vert\alpha\Phi_{AB}+\beta\left(  \frac{I_{AB}-\Phi_{AB}%
}{d^{2}-1}\right)  \right)  \nonumber\\
&  =\inf_{(\alpha,\beta)\in\text{PPT}^{\prime}}\left[  -\log_{2}\inf
_{\Lambda_{AB}\geq0}\left\{
\begin{array}
[c]{c}%
f(\Lambda_{AB},\alpha,\beta):\\
\Lambda_{AB}\leq I_{AB},\\
\operatorname{Tr}[\Lambda_{AB}\Phi_{AB}]=1-\varepsilon
\end{array}
\right\}  \right]  \nonumber\\
&  =-\log_{2}\sup_{(\alpha,\beta)\in\text{PPT}^{\prime}}\inf_{\Lambda_{AB}%
\geq0}\left\{
\begin{array}
[c]{c}%
f(\Lambda_{AB},\alpha,\beta):\\
\Lambda_{AB}\leq I_{AB},\\
\operatorname{Tr}[\Lambda_{AB}\Phi_{AB}]=1-\varepsilon
\end{array}
\right\}  ,
\end{align}
where%
\begin{equation}
f(\Lambda_{AB},\alpha,\beta)\coloneqq \operatorname{Tr}\!\left[  \Lambda_{AB}\left(
\alpha\Phi_{AB}+\beta\left(  \frac{I_{AB}-\Phi_{AB}}{d^{2}-1}\right)  \right)
\right]  .
\label{eq:func-to-simplify}
\end{equation}
Since $\Phi_{AB}$ and $\alpha\Phi_{AB}+\beta\left(  \frac{I_{AB}-\Phi_{AB}%
}{d^{2}-1}\right)  $ are both isotropic in form, it suffices to optimize over
measurement operators $\Lambda_{AB}$ that satisfy the same symmetry, giving
that%
\begin{equation}
\mathcal{T}_{AB}(\Lambda_{AB})=\kappa\Phi_{AB}+\lambda\left(  \frac
{I_{AB}-\Phi_{AB}}{d^{2}-1}\right)  ,
\end{equation}
where we again apply \eqref{eq:result-of-twirl}. The conditions on $\kappa$
and $\lambda$ such that $\mathcal{T}_{AB}(\Lambda_{AB})$ is a measurement
operator are that $\kappa,\lambda\geq0$ and $\kappa,\frac{\lambda}{d^{2}%
-1}\leq1$. Then we find that%
\begin{equation}
\operatorname{Tr}[\Lambda_{AB}\Phi_{AB}]=\operatorname{Tr}[\mathcal{T}%
_{AB}(\Lambda_{AB})\Phi_{AB}]=\kappa,
\end{equation}
so that $\kappa=1-\varepsilon$. Plugging into \eqref{eq:func-to-simplify}, we find
that%
\begin{align}
&  \operatorname{Tr}\!\left[  \Lambda_{AB}\left(  \alpha\Phi_{AB}+\beta\left(
\frac{I_{AB}-\Phi_{AB}}{d^{2}-1}\right)  \right)  \right]  \nonumber\\
&  =\operatorname{Tr}\!\left[  \mathcal{T}_{AB}(\Lambda_{AB})\left(
\alpha\Phi_{AB}+\beta\left(  \frac{I_{AB}-\Phi_{AB}}{d^{2}-1}\right)  \right)
\right]  \\
&  =\operatorname{Tr}\!\left[
\begin{array}
[c]{c}%
\left(  \kappa\Phi_{AB}+\lambda\left(  \frac{I_{AB}-\Phi_{AB}}{d^{2}%
-1}\right)  \right)  \times\\
\left(  \alpha\Phi_{AB}+\beta\left(  \frac{I_{AB}-\Phi_{AB}}{d^{2}-1}\right)
\right)
\end{array}
\right]  \\
&  =\kappa\alpha+\frac{\lambda\beta}{d^{2}-1}=\left(  1-\varepsilon\right)
\alpha+\frac{\lambda\beta}{d^{2}-1}.
\end{align}
Thus, we find that%
\begin{align}
&  \inf_{\sigma_{AB}\in\text{PPT}^{\prime}(A:B)}D_{H}^{\varepsilon}(\rho
_{AB}\Vert\sigma_{AB})\nonumber\\
&  =-\log_{2}\sup_{(\alpha,\beta)\in\text{PPT}^{\prime}}\inf_{\lambda}\left\{
\begin{array}
[c]{c}%
\left(  1-\varepsilon\right)  \alpha+\frac{\lambda\beta}{d^{2}-1}\\
:0\leq\lambda\leq d^{2}-1
\end{array}
\right\}  \\
&  =-\log_{2}\sup_{(\alpha,\beta)\in\text{PPT}^{\prime}}\left(  1-\varepsilon
\right)  \alpha\\
&  =-\log_{2}\left(  1-\varepsilon\right)  \frac{1}{d}=\log_{2}d+\log
_{2}\!\left(  \frac{1}{1-\varepsilon}\right)  .
\end{align}
This concludes the proof.
\end{IEEEproof}

\section{Second law of entanglement dynamics in the asymptotic regime}

As a consequence of Theorem~\ref{thm:main-result}, it follows that the
distillable entanglement does not exceed the entanglement cost (which we
recalled in the introduction is often argued based on physical grounds). To
see this, let us first define the distillable entanglement and entanglement
cost of a bipartite state $\rho_{AB}$. The distillable entanglement is defined
as%
\begin{equation}
E_{D}(A;B)_{\rho}\coloneqq \inf_{\varepsilon\in\left(  0,1\right)  }\liminf
_{n\rightarrow\infty}\frac{1}{n}E_{D}^{\varepsilon}(A^{n};B^{n})_{\rho
^{\otimes n}},\label{def:dist-ent-asymp}%
\end{equation}
and the entanglement cost as%
\begin{equation}
E_{C}(A;B)_{\rho}\coloneqq \sup_{\varepsilon\in\left(  0,1\right)  }\limsup
_{n\rightarrow\infty}\frac{1}{n}E_{C}^{\varepsilon}(A^{n};B^{n})_{\rho
^{\otimes n}}.\label{def:ent-cost-asymp}%
\end{equation}
In words, the distillable entanglement $E_{D}(A;B)_{\rho}$ is equal to the
largest rate at which ebits can be extracted approximately from $\rho_{AB}^{\otimes n}$ by
means of LOCC, such that the error converges to zero in the limit
$n\rightarrow\infty$, and the entanglement cost $E_{C}(A;B)_{\rho}$ is the
smallest rate at which ebits are needed to generate $\rho_{AB}^{\otimes n}$ approximately  by
means of LOCC, such that the error converges to zero in the limit
$n\rightarrow\infty$. The values of $E_{D}(A;B)_{\rho}$ and $E_{C}(A;B)_{\rho
}$ in \eqref{def:dist-ent-asymp}\ and \eqref{def:ent-cost-asymp},
respectively,\ are unchanged by optimizing over $\varepsilon\in(0,c)$ for
$c\in(0,1)$, due to the monotonicity fact stated after \eqref{def:one-shot-ent-cost}.

\begin{corollary}
For every bipartite state $\rho_{AB}$, the following inequality holds%
\begin{equation}
E_{D}(A;B)_{\rho}\leq E_{C}(A;B)_{\rho}.\label{eq:ed-less-than-ec}%
\end{equation}

\end{corollary}

\begin{IEEEproof}
Using the inequality from Theorem~\ref{thm:main-result}, we find for
$\varepsilon\in\left(  0,1/2\right)  $ that%
\begin{align}
&  \liminf_{n\rightarrow\infty}\frac{1}{n}E_{D}^{\varepsilon}(A^{n}%
;B^{n})_{\rho^{\otimes n}}\leq\limsup_{n\rightarrow\infty}\frac{1}{n}%
E_{D}^{\varepsilon}(A^{n};B^{n})_{\rho^{\otimes n}}\nonumber\\
&  \leq\limsup_{n\rightarrow\infty}\frac{1}{n}\left[  E_{C}^{\varepsilon
}(A^{n};B^{n})_{\rho^{\otimes n}}+\log_{2}\!\left(  \frac{1}{1-4\varepsilon
\left(  1-\varepsilon\right)  }\right)  \right]  \nonumber\\
&  =\limsup_{n\rightarrow\infty}\frac{1}{n}E_{C}^{\varepsilon}(A^{n}%
;B^{n})_{\rho^{\otimes n}}.
\end{align}
Taking the infimum over $\varepsilon\in(0,1/2)$ on the left and the supremum
over $\varepsilon\in(0,1/2)$ on the right, we conclude the inequality in \eqref{eq:ed-less-than-ec}.
\end{IEEEproof}

The strong converse distillable entanglement and strong converse entanglement
cost are defined respectively as follows:%
\begin{align}
\widetilde{E}_{D}(A;B)_{\rho} &  \coloneqq \sup_{\varepsilon\in\left(  0,1\right)
}\limsup_{n\rightarrow\infty}\frac{1}{n}E_{D}^{\varepsilon}(A^{n};B^{n}%
)_{\rho^{\otimes n}},\\
\widetilde{E}_{C}(A;B)_{\rho} &  \coloneqq \inf_{\varepsilon\in\left(  0,1\right)
}\liminf_{n\rightarrow\infty}\frac{1}{n}E_{C}^{\varepsilon}(A^n;B^n)_{\rho
^{\otimes n}}.
\end{align}
The inequalities
$
E_{D}(A;B)_{\rho}\leq\widetilde{E}_{D}(A;B)_{\rho}$ and $\widetilde{E}
_{C}(A;B)_{\rho}\leq E_{C}(A;B)_{\rho}
$
are an immediate consequence of definitions. However, it is not clear how to
use the inequality from Theorem~\ref{thm:main-result} to arrive at a similar
statement for the strong converse quantities. That is, the following remains
an open question:%
\begin{equation}
\widetilde{E}_{D}(A;B)_{\rho}\overset{?}{\leq}\widetilde{E}_{C}(A;B)_{\rho}.
\end{equation}

\section{Results for normalized trace distance error}

\label{sec:trace-dist-err}

We can extend the results here to the case when errors are measured by
normalized trace distance, rather than fidelity. Let us define one-shot
distillable entanglement and one-shot entanglement cost using this modified
notion of error:%
\begin{multline}
E_{D}^{\varepsilon,T}(A;B)_{\rho}\coloneqq \\
\sup_{\substack{d\in\mathbb{N},\\\mathcal{L}\in\text{LOCC}}}\left\{  \log
_{2}d:\frac{1}{2}\left\Vert \mathcal{L}_{AB\rightarrow\hat{A}\hat{B}}%
(\rho_{AB})-\Phi_{\hat{A}\hat{B}}^{d}\right\Vert _{1}\leq\varepsilon\right\}
,
\end{multline}%
\begin{multline}
E_{C}^{\varepsilon,T}(A;B)_{\rho}\coloneqq \\
\inf_{\substack{d\in\mathbb{N},\\\mathcal{L}\in\text{LOCC}}}\left\{  \log
_{2}d:\frac{1}{2}\left\Vert \mathcal{L}_{\hat{A}\hat{B}\rightarrow AB}%
(\Phi_{\hat{A}\hat{B}}^{d})-\rho_{AB}\right\Vert _{1}\leq\varepsilon\right\}
.
\end{multline}

\begin{theorem}
Let $\rho_{AB}$ be a bipartite state, and let $\varepsilon_{1},\varepsilon
_{2}\in\left[  0,1\right]  $ be such that $\varepsilon_{1}+\varepsilon_{2}<1$.
Then%
\begin{equation}
E_{D}^{\varepsilon_{2},T}(A;B)_{\rho}\leq E_{C}^{\varepsilon_{1},T}%
(A;B)_{\rho}+\log_{2}\!\left(  \frac{1}{1-\varepsilon_{1}-\varepsilon_{2}%
}\right)  .
\end{equation}

\end{theorem}

\begin{IEEEproof}
The proof idea is essentially the same as that for
Theorem~\ref{thm:main-result}, but we substitute the first part of its proof
with the following bound:%
\begin{equation}
E_{D}^{\varepsilon,T}(A;B)_{\rho}\leq R_{H}^{\varepsilon}(A;B)_{\rho}%
\quad\forall\varepsilon\in\lbrack0,1],\label{eq:trace-dist-ent-dist-rains}%
\end{equation}
and the last part with the triangle inequality for the normalized trace
distance. The inequality in \eqref{eq:trace-dist-ent-dist-rains} follows from
the fact that%
\begin{equation}
\frac{1}{2}\left\Vert \Phi_{AB}-\omega_{AB}\right\Vert _{1}\leq\varepsilon
\quad\Rightarrow\quad\operatorname{Tr}[\Phi_{AB}\omega_{AB}]\geq
1-\varepsilon,\label{eq:trace-dist-imply-succ-meas}%
\end{equation}
along with Proposition~8.6 and Theorem~8.7 of \cite{KW20book}. To see
\eqref{eq:trace-dist-imply-succ-meas}, consider that applying the measurement
channel $(\cdot)\rightarrow\operatorname{Tr}[\Phi_{AB}(\cdot)]|1\rangle
\langle1|+\operatorname{Tr}[(I_{AB}-\Phi_{AB})(\cdot)]|0\rangle\!\langle0|$\ and
the data-processing inequality for trace distance implies that%
\begin{align}
\varepsilon & \geq\frac{1}{2}\left\Vert \Phi_{AB}-\omega_{AB}\right\Vert
_{1}\\
& =\frac{1}{2}\left\Vert
\begin{array}
[c]{c}%
|1\rangle\!\langle1|-\operatorname{Tr}[\Phi_{AB}\omega_{AB}]|1\rangle\!\langle1|\\
-\operatorname{Tr}[(I_{AB}-\Phi_{AB})\omega_{AB}]|0\rangle\!\langle0|
\end{array}
\right\Vert _{1}\\
& =\frac{1}{2}\left\Vert \left(  1-\operatorname{Tr}[\Phi_{AB}\omega
_{AB}]\right)  \left(  |1\rangle\!\langle1|-|0\rangle\!\langle0|\right)
\right\Vert _{1}\\
& =1-\operatorname{Tr}[\Phi_{AB}\omega_{AB}].
\end{align}
This concludes the proof.
\end{IEEEproof}

\section{Discussion}

The result of Theorem~\ref{thm:main-result}\ applies far more generally to the
resource theory of entanglement for bipartite channels
\cite{BHLS03,BDW18,DBW20,BDWW19,GS19}. This follows because the proof of
Theorem~\ref{thm:main-result}\ relies on the sequence of transformations in
\eqref{eq:main-argument-conceptual}. By substituting the state $\rho_{AB}$
there with a bipartite channel, or $n$ sequential uses of it, we conclude that
the same bound holds with $E_{D}^{\varepsilon_{2}}(A;B)_{\rho}$ replaced by
the one-shot distillable entanglement of a bipartite channel and
$E_{C}^{\varepsilon_{1}}(A;B)_{\rho}$ replaced by the one-shot entanglement
cost of the same bipartite channel. Alternatively, these could be replaced by
the $n$-shot distillable entanglement and $n$-shot entanglement cost,
respectively, defined in the sequential way outlined in \cite{BDWW19,GS19}.

More generally, one can extend the reasoning here to arbitrary quantum
resource theories \cite{CG18} (in fact the method used here is the same
conceptually as that used to arrive at Eq.~(51)\ of \cite{Wang2019b} and
Eq.~(50) of \cite{Wang2019a}). The main ingredients needed are a golden-unit
resource like the maximally entangled state, a bound on one-shot distillable
resource like the $\varepsilon$-Rains relative entropy, and an exact
evaluation of the golden-unit resource for this bound.

\section*{Acknowledgment}

I acknowledge discussions with Francesco Buscemi, Siddhartha Das, Nilanjana
Datta, Sumeet Khatri, and Xin Wang, and I acknowledge support from NSF\ Grant No.~1907615.

\bibliographystyle{IEEEtran}
\bibliography{Ref}

\begin{thebibliography}{10}
\providecommand{\url}[1]{#1}
\csname url@samestyle\endcsname
\providecommand{\newblock}{\relax}
\providecommand{\bibinfo}[2]{#2}
\providecommand{\BIBentrySTDinterwordspacing}{\spaceskip=0pt\relax}
\providecommand{\BIBentryALTinterwordstretchfactor}{4}
\providecommand{\BIBentryALTinterwordspacing}{\spaceskip=\fontdimen2\font plus
\BIBentryALTinterwordstretchfactor\fontdimen3\font minus
  \fontdimen4\font\relax}
\providecommand{\BIBforeignlanguage}[2]{{%
\expandafter\ifx\csname l@#1\endcsname\relax
\typeout{** WARNING: IEEEtran.bst: No hyphenation pattern has been}%
\typeout{** loaded for the language `#1'. Using the pattern for}%
\typeout{** the default language instead.}%
\else
\language=\csname l@#1\endcsname
\fi
#2}}
\providecommand{\BIBdecl}{\relax}
\BIBdecl

\bibitem{BBC+93}
C.~H. Bennett, G.~Brassard, C.~Cr\'epeau, R.~Jozsa, A.~Peres, and W.~K.
  Wootters, ``Teleporting an unknown quantum state via dual classical and
  {Einstein-Podolsky-Rosen} channels,'' \emph{Physical Review Letters},
  vol.~70, no.~13, pp. 1895--1899, March 1993.

\bibitem{PhysRevLett.69.2881}
C.~H. Bennett and S.~J. Wiesner, ``Communication via one- and two-particle
  operators on {Einstein-Podolsky-Rosen} states,'' \emph{Physical Review
  Letters}, vol.~69, no.~20, pp. 2881--2884, November 1992.

\bibitem{E91}
A.~K. Ekert, ``Quantum cryptography based on {Bell}'s theorem,'' \emph{Physical
  Review Letters}, vol.~67, no.~6, pp. 661--663, August 1991.

\bibitem{BBPSSW96EPP}
C.~H. Bennett, G.~Brassard, S.~Popescu, B.~Schumacher, J.~A. Smolin, and W.~K.
  Wootters, ``Purification of noisy entanglement and faithful teleportation via
  noisy channels,'' \emph{Physical Review Letters}, vol.~76, no.~5, pp.
  722--725, January 1996, arXiv:quant-ph/9511027.

\bibitem{BDSW96}
C.~H. Bennett, D.~P. DiVincenzo, J.~A. Smolin, and W.~K. Wootters,
  ``Mixed-state entanglement and quantum error correction,'' \emph{Physical
  Review A}, vol.~54, no.~5, pp. 3824--3851, November 1996,
  arXiv:quant-ph/9604024.

\bibitem{HHHH09}
R.~Horodecki, P.~Horodecki, M.~Horodecki, and K.~Horodecki, ``Quantum
  entanglement,'' \emph{Reviews of Modern Physics}, vol.~81, no.~2, pp.
  865--942, June 2009, arXiv:quant-ph/0702225.

\bibitem{PV07}
M.~B. Plenio and S.~Virmani, ``An introduction to entanglement measures,''
  \emph{Quantum Information \& Compution}, vol.~7, no.~1, pp. 1--51, January
  2007, arXiv:quant-ph/0504163.

\bibitem{W18thesis}
X.~Wang, ``Semidefinite optimization for quantum information,'' Ph.D.
  dissertation, University of Technology Sydney, Centre for Quantum Software
  and Information, Faculty of Engineering and Information Technology, July
  2018, \url{http://hdl.handle.net/10453/127996}.

\bibitem{KW20book}
S.~Khatri and M.~M. Wilde, \emph{Principles of Quantum Communication Theory: A
  Modern Approach}, Nov. 2020, arXiv:2011.04672v1.

\bibitem{HHT01}
P.~M. Hayden, M.~Horodecki, and B.~M. Terhal, ``The asymptotic entanglement
  cost of preparing a quantum state,'' \emph{Journal of Physics A: Mathematical
  and General}, vol.~34, no.~35, p. 6891, Sep. 2001, arXiv:quant-ph/0008134.

\bibitem{WCP11}
M.~M. Wolf, T.~S. Cubitt, and D.~Perez-Garcia, ``Are problems in quantum
  information theory (un)decidable?'' Nov. 2011, arXiv:1111.5425.

\bibitem{PhysRevA.65.012323}
G.~Vidal and J.~I. Cirac, ``Irreversibility in asymptotic manipulations of a
  distillable entangled state,'' \emph{Physical Review A}, vol.~65, no.~1, p.
  012323, Dec. 2001.

\bibitem{Wat18}
J.~Watrous, \emph{The Theory of Quantum Information}.\hskip 1em plus 0.5em
  minus 0.4em\relax Cambridge University Press, 2018.

\bibitem{BP08}
F.~G. Brandao and M.~B. Plenio, ``Entanglement theory and the second law of
  thermodynamics,'' \emph{Nature Physics}, vol.~4, pp. 873–--877, October
  2008, arXiv:0810.2319.

\bibitem{BP2010}
F.~G. S.~L. Brand{\~a}o and M.~B. Plenio, ``A reversible theory of entanglement
  and its relation to the second law,'' \emph{Communications in Mathematical
  Physics}, vol. 295, no.~3, pp. 829--851, May 2010, arXiv:0710.5827.

\bibitem{BD10a}
F.~Buscemi and N.~Datta, ``Distilling entanglement from arbitrary resources,''
  \emph{Journal of Mathematical Physics}, vol.~51, no.~10, p. 102201, October
  2010, arXiv:1006.1896.

\bibitem{BD11}
------, ``Entanglement cost in practical scenarios,'' \emph{Physical Review
  Letters}, vol. 106, no.~13, p. 130503, Mar. 2011, arXiv:0906.3698.

\bibitem{TWW17}
M.~Tomamichel, M.~M. Wilde, and A.~Winter, ``Strong converse rates for quantum
  communication,'' \emph{{IEEE} Transactions on Information Theory}, vol.~63,
  no.~1, pp. 715--727, January 2017, arXiv:1406.2946.

\bibitem{TBR15}
M.~Tomamichel, M.~Berta, and J.~M. Renes, ``Quantum coding with finite
  resources,'' \emph{Nature Communications}, vol.~7, p. 11419, May 2016,
  arXiv:1504.04617.

\bibitem{FWTD19}
K.~Fang, X.~Wang, M.~Tomamichel, and R.~Duan, ``Non-asymptotic entanglement
  distillation,'' \emph{IEEE Transactions on Information Theory}, vol.~65,
  no.~10, pp. 6454--6465, Oct. 2019, arXiv:1706.06221.

\bibitem{BHLS03}
C.~H. Bennett, A.~W. Harrow, D.~W. Leung, and J.~A. Smolin, ``On the capacities
  of bipartite {Hamiltonians} and unitary gates,'' \emph{IEEE Transactions on
  Information Theory}, vol.~49, no.~8, pp. 1895--1911, August 2003,
  arXiv:quant-ph/0205057.

\bibitem{BDW18}
S.~B\"auml, S.~Das, and M.~M. Wilde, ``Fundamental limits on the capacities of
  bipartite quantum interactions,'' \emph{Physical Review Letters}, vol. 121,
  no.~25, p. 250504, Dec. 2018, arXiv:1812.08223.

\bibitem{DBW20}
S.~Das, S.~B\"auml, and M.~M. Wilde, ``Entanglement and secret-key-agreement
  capacities of bipartite quantum interactions and read-only memory devices,''
  \emph{Physical Review A}, vol. 101, no.~1, p. 012344, Jan. 2020,
  arXiv:1712.00827.

\bibitem{BDWW19}
S.~B\"auml, S.~Das, X.~Wang, and M.~M. Wilde, ``Resource theory of entanglement
  for bipartite quantum channels,'' July 2019, arXiv:1907.04181.

\bibitem{GS19}
G.~Gour and C.~M. Scandolo, ``The entanglement of a bipartite channel,'' Jul.
  2019, arXiv:1907.02552.

\bibitem{CLM+14}
E.~Chitambar, D.~Leung, L.~Man{\v{c}}inska, M.~Ozols, and A.~Winter,
  ``Everything you always wanted to know about {LOCC} (but were afraid to
  ask),'' \emph{Communications in Mathematical Physics}, vol. 328, no.~1, pp.
  303--326, May 2014, arXiv:1210.4583.

\bibitem{PhysRevA.59.1070}
C.~H. Bennett, D.~P. DiVincenzo, C.~A. Fuchs, T.~Mor, E.~Rains, P.~W. Shor,
  J.~A. Smolin, and W.~K. Wootters, ``Quantum nonlocality without
  entanglement,'' \emph{Physical Review A}, vol.~59, no.~2, pp. 1070--1091,
  February 1999.

\bibitem{Uhl76}
A.~Uhlmann, ``The `transition probability' in the state space of a *-algebra,''
  \emph{Reports on Mathematical Physics}, vol.~9, no.~2, pp. 273--279, April
  1976.

\bibitem{R02}
A.~E. Rastegin, ``Relative error of state-dependent cloning,'' \emph{Physical
  Review A}, vol.~66, no.~4, p. 042304, October 2002.

\bibitem{R03}
------, ``A lower bound on the relative error of mixed-state cloning and
  related operations,'' \emph{Journal of Optics B: Quantum and Semiclassical
  Optics}, vol.~5, no.~6, p. S647, December 2003, arXiv:quant-ph/0208159.

\bibitem{GLN04}
A.~Gilchrist, N.~K. Langford, and M.~A. Nielsen, ``Distance measures to compare
  real and ideal quantum processes,'' \emph{Physical Review A}, vol.~71, no.~6,
  p. 062310, June 2005, arXiv:quant-ph/0408063.

\bibitem{R06}
A.~E. Rastegin, ``Sine distance for quantum states,'' February 2006,
  arXiv:quant-ph/0602112.

\bibitem{BD10}
F.~Buscemi and N.~Datta, ``The quantum capacity of channels with arbitrarily
  correlated noise,'' \emph{IEEE Transactions on Information Theory}, vol.~56,
  no.~3, pp. 1447--1460, March 2010, arXiv:0902.0158.

\bibitem{BD11ieee}
F.~G. S.~L. Brandao and N.~Datta, ``One-shot rates for entanglement
  manipulation under non-entangling maps,'' \emph{IEEE Transactions on
  Information Theory}, vol.~57, no.~3, pp. 1754--1760, March 2011,
  arXiv:0905.2673.

\bibitem{WR12}
L.~Wang and R.~Renner, ``One-shot classical-quantum capacity and hypothesis
  testing,'' \emph{Physical Review Letters}, vol. 108, no.~20, p. 200501, 2012,
  arXiv:1007.5456.

\bibitem{AdMVW02}
K.~Audenaert, B.~De~Moor, K.~G.~H. Vollbrecht, and R.~F. Werner, ``Asymptotic
  relative entropy of entanglement for orthogonally invariant states,''
  \emph{Physical Review~A}, vol.~66, no.~3, p. 032310, September 2002,
  arXiv:quant-ph/0204143.

\bibitem{KW17a}
E.~Kaur and M.~M. Wilde, ``Upper bounds on secret key agreement over lossy
  thermal bosonic channels,'' \emph{Physical Review A}, vol.~96, no.~6, p.
  062318, Dec. 2017, arXiv:1706.04590.

\bibitem{R99}
E.~M. Rains, ``Bound on distillable entanglement,'' \emph{Physical Review A},
  vol.~60, pp. 179--184, July 1999.

\bibitem{R01}
------, ``A semidefinite program for distillable entanglement,'' \emph{IEEE
  Transactions on Information Theory}, vol.~47, no.~7, pp. 2921--2933, November
  2001, arXiv:quant-ph/0008047.

\bibitem{CG18}
E.~Chitambar and G.~Gour, ``Quantum resource theories,'' \emph{Reviews of
  Modern Physics}, vol.~91, no.~2, p. 025001, Apr. 2019, arXiv:1806.06107.

\bibitem{Wang2019b}
X.~Wang and M.~M. Wilde, ``Resource theory of asymmetric distinguishability,''
  \emph{Physical Review Research}, vol.~1, no.~3, p. 033170, Dec. 2019,
  arXiv:1905.11629.

\bibitem{Wang2019a}
------, ``Resource theory of asymmetric distinguishability for quantum
  channels,'' \emph{Physical Review Research}, vol.~1, no.~3, p. 033169, Dec.
  2019, arXiv:1907.06306.

\end{thebibliography}

\end{document}